# The impact of unproductive and top professors on overall university research performance[1]

*Giovanni Abramo[a,b,\*], Tindaro Cicero[b], Ciriaco Andrea D'Angelo[b]*

[a] Institute for System Analysis and Computer Science (IASI-CNR)
National Research Council of Italy

[b] Laboratory for Studies of Research and Technology Transfer
School of Engineering, Department of Management
University of Rome "Tor Vergata"

**Abstract**

Unlike competitive higher education systems, non-competitive systems show relatively uniform distributions of top professors and low performers among universities. In this study, we examine the impact of unproductive and top faculty members on overall research performance of the university they belong to. Furthermore, we analyze the potential relationship between research productivity of a university and the indexes of concentration of unproductive and top professors. Research performance is evaluated using a bibliometric approach, through publications indexed on the *Web of Science* between 2004 and 2008. The set analyzed consists of all Italian universities active in the hard sciences.

---

\* **Corresponding author:** Dipartimento di Ingegneria dell'Impresa, Università degli Studi di Roma "Tor Vergata", Via del Politecnico 1, 00133 Rome - ITALY, tel/fax +39 06 72597362, giovanni.abramo@uniroma2.it

# 1. Introduction

National systems of higher education differ in intensity of competition among universities due to general differences in funding mechanisms, levels of institutional autonomy, and culture. The "Anglo-Saxon" tradition, particularly in the United States, has led to competitive university systems, with emergence of world-class research-intensive universities and notable inequality in performance among institutions. In the area of research performance, such levels of heterogeneity are confirmed by relatively high values of the Gini concentration index, although referred to the gross performance indicator of publications per researcher (Halffman and Leydesdorff, 2010). In competitive systems, in fact, there is a concentration of top professors in elite universities and low performers in the lower prestige institutions.

In Northern Europe, the development of New Public Management in the academic sector, with emphasis on quasi-market competition, efficiency and performance audit practices, will likely lead to an overall increase in performance, together with greater differentiation among universities. Expectations are also that, in the coming years, government funding will be distributed in an increasingly less uniform manner (Horta et al., 2008). In southern Europe, higher education systems are generally composed of public universities with relatively low autonomy, and are often characterized by weak overall performance with little differentiation among institutions (Van der Ploeg and Veugelers, 2008). In Italy, as a case in point, the variability of research productivity among universities (standardized citations per researcher in the same field) results as being much lower than that within individual institutions (Abramo et al., 2012a). Essentially, both top and bottom performing professors are found in the same universities.

In a growing number of countries, universities are ever more subject to research performance evaluations and ranking assessments (Osterloh and Frey, 2008; Leydesdorff, 2008; Hicks, 2011). However, when national research assessment exercises inform the allocation of public funds to individual universities in non-competitive, undifferentiated systems, the paradox could develop that low performing professors in higher-performing universities would receive more funds than higher performers in low-performing universities. In this work we examine the impact of unproductive[2] faculty members in the classification of individual universities in undifferentiated systems, comparing the research productivity rankings of universities both including and excluding unproductive professors from the calculations. For each university, we also measure the levels of concentration of unproductive and top professors, to verify any negative correlation. Finally, we attempt to identify a relation between research productivity of a university and indexes of concentration of unproductive and of top professors. The analyses refer to the Italian university system, which offers a good example of scarce differentiation in research productivity.

In the next section we describe the principle characteristics of the higher education system seen in Italy. In Section 3 we explain the dataset and methodology for the indicators of productivity used in the analysis. Section 4 provides the results of the analysis and the final section offers the study conclusions and the authors' considerations.

---

[2] Unproductive refers to the research activity of faculty members. We define a professor unproductive if his research productivity, as defined in section 3, is nil.

## 2. The Italian higher education system

The Italian Ministry of Education, Universities and Research (MIUR) officially recognizes a total of 96 universities, with authority to issue legally-recognized degrees. Twenty-nine of these are very small private special-focus universities, of which 13 offer only e-learning. Sixty-seven are public and generally multi-disciplinary universities, scattered throughout the nation, some having a number of branches in smaller cities. Six of them are *Scuole Superiori* (Schools for Higher Studies), specifically devoted to highly talented students, with very small faculties and tightly limited enrollment numbers per degree program. In Italy, 94.9% of faculty are employed in public universities (0.5% in Scuole Superiori) and 5.1% are in private universities.

The Italian higher education system is a long-standing, classic example of a public and highly centralized governance structure, with low levels of autonomy at the university level and a very strong role played by the central state. To date, the most significant intervention for liberalization has been Law 168 of 1989, intended to grant increased autonomy and responsibilities to the universities. In particular, Articles 6 and 7 of this law are intended to enact Article 37 of the national Constitution, establishing the fundamentals of university autonomy in teaching, research, financing and accounting, and directing that individual institutions appropriately establish an autonomous organizational framework including their own charters and regulations. Law 537 (Article 5) of 1993 and Decree 168 of 1996 introduced substantial changes in central financing for universities, specifically in relation to overall amounts, re-equilibration among institutions, freedom and responsibility in allocating expenses, involvement in fund-raising, and freedom to apply tuition fees provided that these do not exceed 20% of total government funding. Law 537 had two objectives: to increase university involvement in overall decision-making relating to use of resources, and to encourage individual institutions to operate on the market and reach their own economic and financial equilibrium.

In keeping with the Humboldt model, there are no "teaching-only" universities in Italy, as all professors are required to carry out both research and teaching. National regulations establish that each faculty member must allocate a minimum of 350 hours per year to teaching. At the close of 2011, there were around 59,000 faculty members in Italy (full, associate and assistant professors) and a roughly equal number of technical-administrative staff. All new personnel enter the university system through public examinations and career advancement can only proceed by further public examinations. Salaries are regulated at the centralized level and are calculated according to role (administrative, technical, or professorial), rank within role (for example: assistant, associate or full professor) and seniority. None of a professor's salary depends on merit: salaries increase annually according to rules set by government. Moreover, as in all Italian public administration, dismissal of an employee for lack of productivity is unheard of.

The whole of these conditions create an environment and a culture that are completely non-competitive, yet flourishing with favoritism and other opportunistic behaviors that are dysfunctional to the social and economic roles of the higher education system. The overall result is a system of universities that are almost completely undifferentiated for quality and prestige, with the exception of the tiny Scuole Superiori and a very small number of the private special-focus universities. The system is thus unable to attract significant foreign faculty or students. The numbers are negligible:

foreign students are 3% of the total, compared to the OECD average of 8.5%, and only 2.3% of actual graduates are foreigners; only 1.8% of research staff are foreign nationals. This is a system where every university has some share of top professors, flanked by another share of absolute non-producers. Over the 2004-2008 period, 6,640 (16.8%) of the 39,512 hard sciences professors did not publish any scientific articles in the journals indexed by the Thomson Reuters' *Web of Science* (WoS). Another 3,070 professors (7.8%) did achieve publication, but their work was never cited. This means that 9,710 individuals (24.6%) had no impact on scientific progress. An almost equal 23.0% of professors alone produced 77% of the overall scientific advancement. The problem is that this 23% of faculty is not concentrated in a limited number of universities, but is instead dispersed more or less uniformly among all Italian universities, along with the unproductive ones, so that no single institution reaches the critical mass of excellence necessary to develop as an elite university and compete at the international level (Abramo et al., 2012a). Given the situation, it can be no surprise that no Italian university ranks above 150th position in any of the yearly world universities rankings (THES, 2012; SJTU, 2011; QS, 2011).

## 3. Productivity rankings at individual and university levels: methodology and dataset

Research activity is a production process in which the inputs consist of human, tangible (scientific instruments, materials, etc.) and intangible (accumulated knowledge, social networks, etc.) resources, and where outputs have a complex character of both tangible nature (publications, patents, conference presentations, databases, protocols, etc.) and intangible nature (tacit knowledge, consulting activity, etc.). The new-knowledge production function has therefore a multi-input and multi-output character. The principal efficiency indicator of any production system is labor productivity. To calculate it one needs adopt a few simplifications and assumptions. It has been shown (Moed, 2005) that in the hard sciences, including life sciences, the prevalent form of codification of research output is the publication in scientific journals. As a proxy of total output in this work we consider only publications (articles, article reviews, and proceeding papers) indexed in the WoS. The other forms of output which we neglect are often followed by publications that describe their content in the scientific arena, so the analysis of publications alone actually avoids a potential double counting.

When measuring labor productivity, if there are differences in the production factors available to each professor then one should normalize by them. Unfortunately relevant data are not available at individual level in Italy. The first assumption then is that resources available to professors within the same field of observation are the same. The second assumption is that the hours devoted to research are more or less the same for all professors. In Italy the above assumptions are acceptable, because in the period of observation core government funding was input oriented, and distributed to satisfy the resource needs of each and every university in function of their size and activities. Furthermore, the hours that each professor has to devote to teaching are established by national regulations and the same for all.

Research projects frequently involve a team of researchers, which shows in co-authorship of publications. Productivity measures then need to account for the fractional contributions of professors to their outputs. In the life science, the position of co-authors in the list reflects the relative contribution to the project and needs to be weighted accordingly. Furthermore, because the intensity of publications varies across fields (Abramo et al., 2008), in order to avoid distortions in productivity rankings, one must compare researchers within the same field. A prerequisite of any research performance assessment free of distortions is then a classification of each researcher in one and only one field. In the Italian university system all professors are classified in one field. To our knowledge, this feature of the Italian higher education system is unique in the world. In the hard sciences, there are 205 such fields (named scientific disciplinary sectors, SDSs[3]), grouped into nine disciplines (named university disciplinary areas, UDAs[4]). Since it has been demonstrated that productivity of full, associate and assistant professors is different (Abramo et al., 2011a), their distribution is not uniform across universities, and academic rank determines differentiation in stipends, we differentiate performance rankings by academic rank.

A very gross way to calculate the average yearly labor research productivity is to simply measure the weighted fractional count of publications per researcher in the period of observation and divide it for the full time equivalent of work in the period. A more sophisticated way to calculate productivity recognizes the fact that publications, embedding the new knowledge produced, have different values. Their value depends on their impact on scientific advancements. As proxy of impact bibliometricians adopt the number of citations for the researchers' publications.

However, comparing researchers' performance by field and academic rank is not enough to avoid distortions in rankings. In fact citation behavior too varies across fields, and it has been shown that it is not unlikely that researchers belonging to a particular scientific field may also publish outside that field (a typical example is statisticians, who may apply theory to medicine, physics, social sciences, etc.). For this reason we standardize the citations for each publication accumulated at June 30, 2009 with respect to the median[5] for the distribution of citations for all the Italian publications of the same year and the same subject category[6].

In formulae, the average yearly productivity at the individual level, *p* is the following:

$$p = \frac{1}{t} \cdot \sum_{i=1}^{N} \frac{c_i}{Me_i} * \frac{1}{s_i} \qquad [1]$$

Where:
t = number of years of work of the researcher in the period of observation
N = number of publications of the researcher in the period of observation.
$c_i$ = citations received by publication *i*;
$Me_i$ = median of the distribution of citations received for all Italian cited-only

---

[3] The complete list is accessible on http://attiministeriali.miur.it/UserFiles/115.htm
[4] Mathematics and computer sciences; physics; chemistry; earth sciences; biology; medicine; agricultural and veterinary sciences; civil engineering; industrial and information engineering.
[5] As frequently observed in literature (Lundberg, 2007), standardization of citations with respect to median value rather than to the average is justified by the fact that distribution of citations is highly skewed in almost all disciplines.
[6] The subject category of a publication corresponds to that of the journal where it is published. For publications in multidisciplinary journals the scaling factor is calculated as a weighted average of the standardized values for each subject category.

publications of the same year and subject category of publication $i$;
$s_i$ = co-authors of publication $i$

In the life sciences, widespread practice is for the authors to indicate the various contributions to the published research by the positioning of the names in the authors list. For life sciences then, when the number of co-authors is higher than two, different weights are given to each co-author according to his/her position in the list and the character of the co-authorship (intra-mural or extra-mural). If first and last authors belong to the same university, 40% of citations are attributed to each of them; the remaining 20% are divided among all other authors. If the first two and last two authors belong to different universities, 30% of citations are attributed to first and last authors; 15% of citations are attributed to second and last author but one; the remaining 10% are divided among all others[7].

Based on the value of $p$ we obtain, for each SDS, a ranking list expressed in percentiles and differentiated by academic rank. Thus the performance of each professor is calculated in each SDS for each academic rank and expressed on a percentile scale of 0-100 (worst to best) for comparison with the performance of all Italian colleagues of the same academic rank and SDS; or as the ratio to the average performance of all Italian colleagues of the same academic rank and SDS. We can exclude, for the Italian case, that productivity ranking lists may be distorted by variable returns to scale, due to different sizes of universities (Abramo et al., 2012b).

Data on faculty of each university and their SDS classification are extracted from the database on Italian university personnel, maintained by the MIUR. The bibliometric dataset used to measure $p$ is extracted from the Italian Observatory of Public Research (ORP), a database developed and maintained by the authors and derived under license from the WoS. Beginning from the raw data of the WoS, and applying a complex algorithm for reconciliation of the author's affiliation and disambiguation of the true identity of the authors, each publication (article, article review and conference proceeding) is attributed to the university professor or professors that produced it (D'Angelo et al., 2011). Thanks to this algorithm, we can produce rankings of research productivity at the individual level, on a national scale.

Average yearly productivity $P$ of an entire university is given by sum of the individuals' productivities, each normalized to the average per academic rank in the SDS, divided by the total number of professors, thus:

$$P = \frac{1}{N} \cdot \sum_{j=1}^{3} \sum_{k=1}^{n} \left( \frac{p_{jk}}{Ap_{jk}} \right) \qquad [2]$$

Where:
N = number of researchers at the university
n = number of SDSs where the university is active
$p$ = researcher productivity
$Ap$ = national average of researcher productivity
j indicates the academic rank
k indicates the SDS

---

[7] The weighting values were assigned following advice from Italian professors in the life sciences. The values could be changed to suit different practices in other national contexts.

To calculate productivity net of unproductive professors, the formula denominator then excludes all professors with nil productivity value over the five years under examination (*N_prod*), thus:

$$P(1) = \frac{1}{N\_prod} \cdot \sum_{j=1}^{3} \sum_{k=1}^{n} \left(\frac{P_{jk}}{Ap_{jk}}\right) \qquad [3]$$

On the basis of the count of professors with nil productivity values, we calculate the index of concentration for unproductive professors, *NR*, for every university. Since the concentration of unproductive professors varies among SDSs and academic ranks, and the distribution of SDSs and their professors is not uniform among universities, we apply the following formula to avoid distortion:

$$NR = \frac{1}{N} \cdot \sum_{k=1}^{n} \frac{\%\,IS_k}{\%AIS_k} * N_k \qquad [4]$$

Where:
N = total number of professors at the university
%IS = ratio of unproductive professors to total professors in the SDS
%AIS = national average ratio of unproductive professors to total professors in the SDS
k indicates the SDS

Finally, we calculate the index of concentration of top professors *TR*, representing the number of professors classified in the top 20%[8] in Italy, for research productivity in their SDS and for their academic rank, relative to the total professors at the university, thus:

$$TR = \frac{top\ 20\%\ scientists}{N} \qquad [5]$$

Table 1 lists the indicators described above with the acronyms that are used in the remainder of the article.

| Indicator | Formula | Description |
|---|---|---|
| P | [2] | Total productivity of the university |
| P(1) | [3] | Total productivity of the university, net of unproductive professors |
| NR | [4] | Index of concentration of unproductive professors per university |
| TR | [5] | Index of concentration of top 20% professors per university |

*Table 1: Summary of indicators used for the analysis*

---

[8] The threshold value could be changed according to preferences. The 20% value is suggested here because the analysis is carried out at the SDS level, and in a few of them the number of professors is not very high.

## 3. Results and analysis

### 3.1. Descriptive analysis

To render WoS-indexed publications a more robust proxy of overall output of a professor, the field of observation is limited to those SDSs where at least 50% of the professors produced at least one publication in the period 2004-2008. There are 184 such SDSs. We further exclude the universities with less than five professors in any specific SDS considered. The field of analysis is thus limited to 39,477 professors from 65 of the total 89 Italian universities.

As a first step, we calculate the descriptive statistics for the distributions of university performance indicators (Table 2).

|                     | P      | P(1)   | NR     | TR     |
|---------------------|--------|--------|--------|--------|
| Observations        | 65     | 65     | 65     | 65     |
| Mean                | 1.066  | 1.393  | 0.981  | 0.214  |
| Std dev.            | 0.419  | 0.422  | 0.301  | 0.084  |
| Minimum             | 0.278  | 0.784  | 0.187  | 0.054  |
| Maximum             | 2.601  | 2.680  | 1.865  | 0.548  |
| Coeff. of variation | 39.274 | 30.289 | 30.699 | 39.143 |
| First quartile      | 0.772  | 1.108  | 0.779  | 0.161  |
| Median              | 1.014  | 1.288  | 0.951  | 0.204  |
| Third quartile      | 1.230  | 1.555  | 1.174  | 0.251  |
| Skewness            | 1.575  | 1.440  | 0.297  | 1.411  |
| Gini coefficient    | 0.198  | 0.150  | 0.170  | 0.202  |

*Table 2: Descriptive statistics for indicators used for the analysis*

The first two columns show the descriptive statistics for the overall university productivity P and productivity excluding unproductive professors, P(1). With unproductive professors excluded, we observe that the performance distribution of the universities tends to be slightly less dispersed around an average value of productivity that is obviously higher, increasing from 1066 to 1.393. With the increased average value there is actually a slight decrease in the coefficient of variation. The Gini coefficient also decreases, going from 0.198 to 1.150.

This slight drop in dispersion shows that the unproductive professors are found in a quite uniform manner in almost all universities. In a competitive system we probably would witness a drastic drop in the dispersion of performance between universities, since the unproductive professors would be concentrated in a subgroup of institutions. Avoiding their inclusion in the productivity measure would permit their universities to notably reduce their performance gap with the best universities.

The box plot in Figure 1 shows the comparison between the two scenarios created.

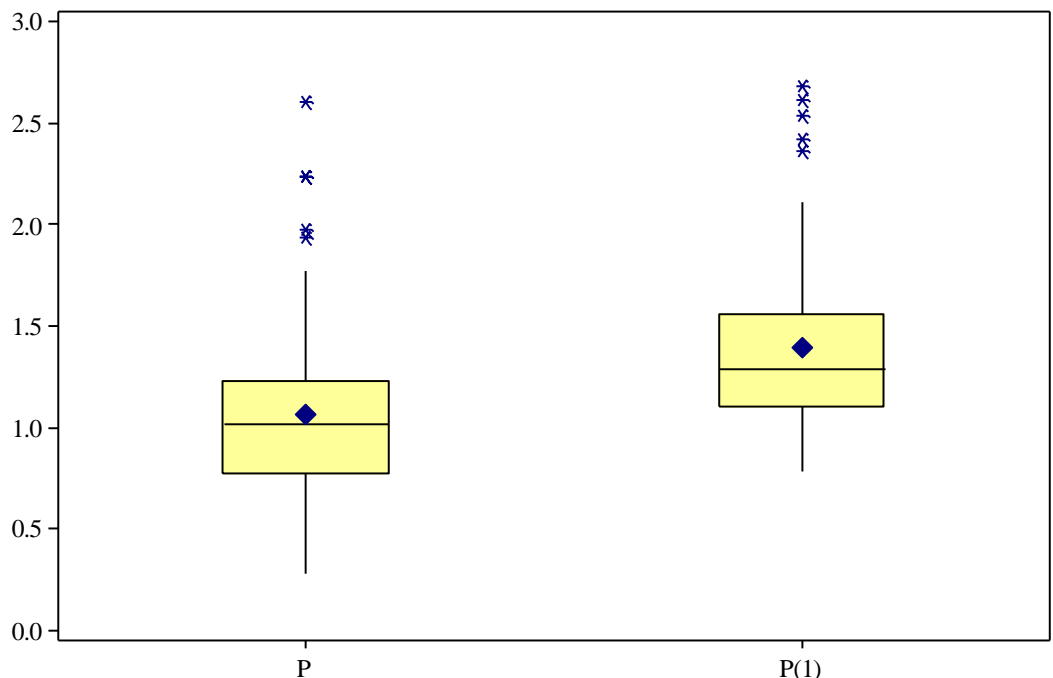

*Figure 1: Comparison between box plots for distributions of university productivity, with and without unproductive professors*

Among the outliers, the first three universities maintain the first three positions under both scenarios since the share of unproductive professors is so low that it only weakly influences performance.

Overall, excluding the unproductive professors leads to 35 out of 65 universities experiencing an increase in productivity of between 25 and 50% (Figure 2), while 23 show an increase of between 0 and 25%.

There are also three anomalous cases: one university experiences a productivity increase of between 80 and 100%, another an increase between 100 and 120%, and a third experiences an increase of an extreme 350%, being specialized in a research area where the significance of bibliometric analysis is at the limit and thus performance is extremely sensitive to the presence or exclusion of the unproductive professors.

In comparing the P and P(1) ranking lists we observe variations in rank for the individual universities. Table 3 presents the descriptive statistics for the distribution of these variations, as absolute and real values.

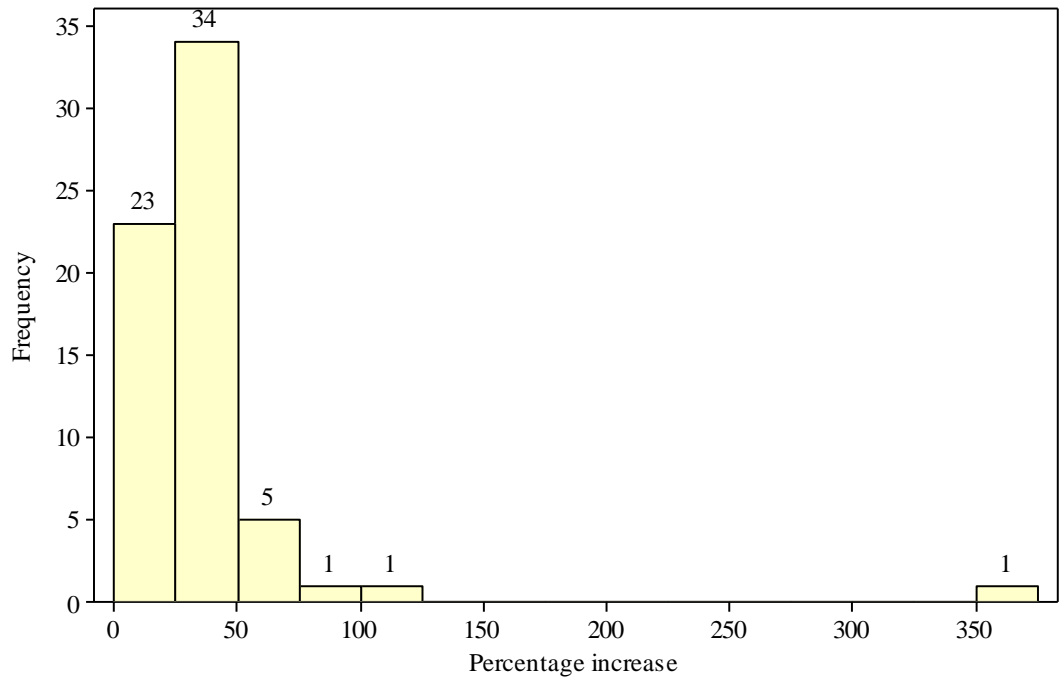

*Figure 2: Distribution of percentage increase of performance from P to P(1)*

| Statistics | Absolute variations | Real value variations |
|---|---|---|
| No variations | 6 | 6 |
| Variations | 59 | 59 |
| Average | 5.323 | 0 |
| Median | 3 | -1 |
| Skewness | 2.341 | 1.869 |
| Std dev. | 5.748 | 8.436 |
| Minimum | 0 | -15 |
| Maximum | 31 | +31 |

*Table 3: Descriptive statistics of absolute and real value variations between rankings by P and by P(1)*

This analysis shows that 59 out of 65 universities change at least one position. An individual university experiences an average absolute rank variation of 5.3 positions, which is greater than that median variation, at 3. For most universities (about 40 out of 65), the variation in rank is between 0 and 4 positions. The average value is particularly influenced by two universities that experience a jump of more than 25 positions in rank and thus determine the strong positive asymmetry (skewness = 2.341) of the absolute variations in rank.

Table 3, column 3 presents the statistics on the variations in rank: this permits us to distinguish the negative from the positive variations. Naturally the distribution of these variations has an average value of nil, since the negative and positive shifts compensate. Still, we note that the maximum positive value of such variations is 31 positions, which is much more than the largest negative shift, at -15, in this case determining a less marked positive asymmetry (skewness=1.869.). In general, the ranks based on the two forms of productivity are quite similar, with the Spearman correlation value being +0.898 (p-value<0.01).

We next classify the universities in four groups based on their P and P(1) values (a common method of applying national research assessment results). We assign values of 4, 3, 2 and 1 corresponding to the first, second, third and fourth quartiles for productivity value. Then we calculate the distributions of the shifts in quartile that a university experiences in moving from ranking by P to that net of unproductive professors, by P(1) (Table 4).

|   |   | P(1) |   |   |   |
|---|---|------|---|---|---|
|   |   | 4    | 3 | 2 | 1 |
| P | 4 | 13   | 3 | 0 | 0 |
|   | 3 | 3    | 8 | 5 | 0 |
|   | 2 | 0    | 5 | 8 | 3 |
|   | 1 | 0    | 0 | 3 | 14|

*Table 4: Analysis of changes in quartile between rankings by P and by P(1)*

No university makes a shift of more than one quartile, thus confirming the similarity between the two ranking lists. Still, if we hypothesize a university financing system in Italy that resembles that of the UK's, where funds are allocated only to universities that occupy the top two performance quartiles as assigned by research assessment exercises, we would see that five universities that would not receive funds on account of unproductive professors, would now receive funds with the removal of unproductive professors in considering their productivity.

Returning to Table 2, the fourth and fifth columns refer to the concentration of unproductive and top professors in the 65 Italian universities. As seen in Figure 3, the distribution of the index of unproductive professors (NR) is more concentrated around its median value (0.981) than the distribution of the tops professor index (TR), which is more asymmetric to the right; 16 universities out of 65 have a TR value between 15% and 20%, 17 have a value between 20% and 25%, and finally 6 universities (9%) have a TR concentration greater than 30%, establishing the right tail of the distribution.

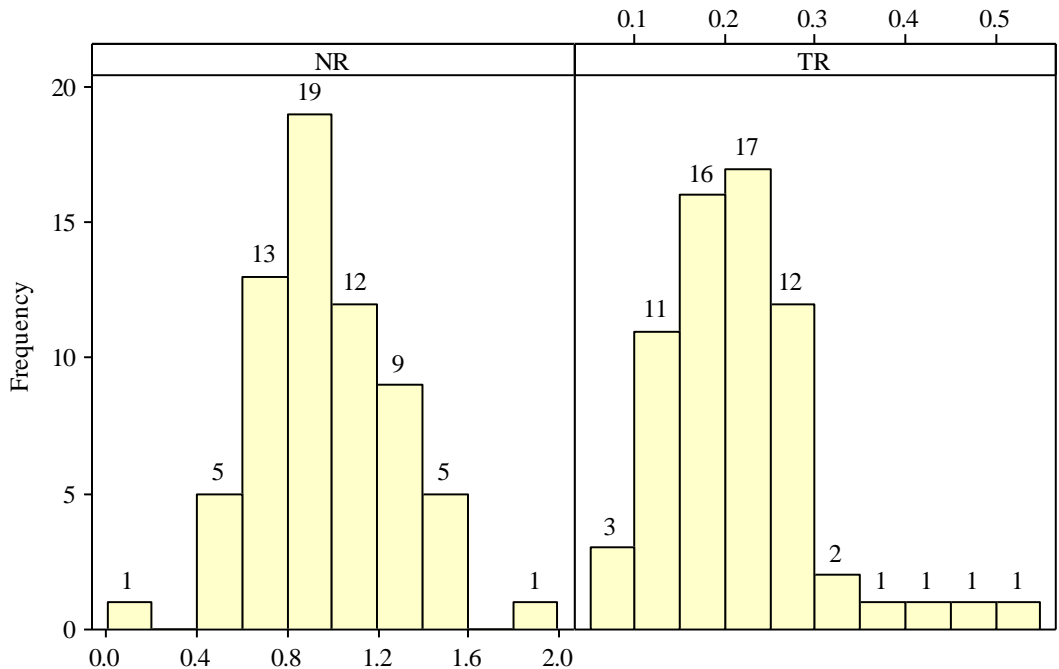

*Figure 3: Distributions of NR and TR among 65 Italian universities*

From this first descriptive analysis we see that unproductive professors are distributed quite equally among Italian universities, while the concentration of top professors is more dispersed. With removal of their unproductive professors, some universities experience significant increases in productivity, but these do not translate into particularly evident shifts in rank, and are never more than a one quartile shift. Observing these various distributions, in the next section we analyze the potential relationship between overall productivity of a university and its rates of unproductive and top professors.

**3.2. Relationship between university productivity and the concentrations of top and unproductive professors**

To evaluate the relationship between the overall productivity of a university and the concentrations of top and unproductive professors, we first examine the correlations between the indicators described above (Table 5).

|    | P       | TR      | NR    |
|----|---------|---------|-------|
| P  | 1.000   |         |       |
| TR | 0.953*  | 1.000   |       |
| NR | -0.720* | -0.662* | 1.000 |

*Table 5: Spearman correlation matrix of variables for regression analysis*
*\* p-value < 0.01*

The Spearman correlation matrix shows that concentration of top professors (TR) is negatively correlated (Spearman $\rho$= -0.662) with that of unproductive professors (NR). In spite of there being a quite uniform distribution of unproductive and top professors and within Italian universities, the ranking lists derived from the TR and NR indicators are inversely correlated: universities that place high in the classification for TR tend to occupy lower positions in the classification for NR, and vice versa, The anomalous cases, different than this tendency, are the universities situated in the upper right and lower left quadrants in Figure 4. The most interesting cases, indicated with larger triangles, are two universities that show totally superimposed values, notably lower than the median, for NR and TR; a third university that shows values notably higher than the median for both indicators, and finally a fourth university that has the highest concentration of top professors, but an NR value around the median.

Table 5 also shows that both the TR and NR indicators are strongly and consistently correlated with productivity P: TR is positively correlated (Spearman $\rho$= +0.953), while NR is negatively correlated (Spearman $\rho$= -0.720). The right side of Figure 5 shows the scatter plot for correlation between productivity and concentration of top professors. We observe that the relationship is linear, with particular concentration of universities around the median of P (1.014). The left side of the same figure correlates productivity and concentration of unproductive professors: in this case the linear relationship is less evident and the values are more dispersed.

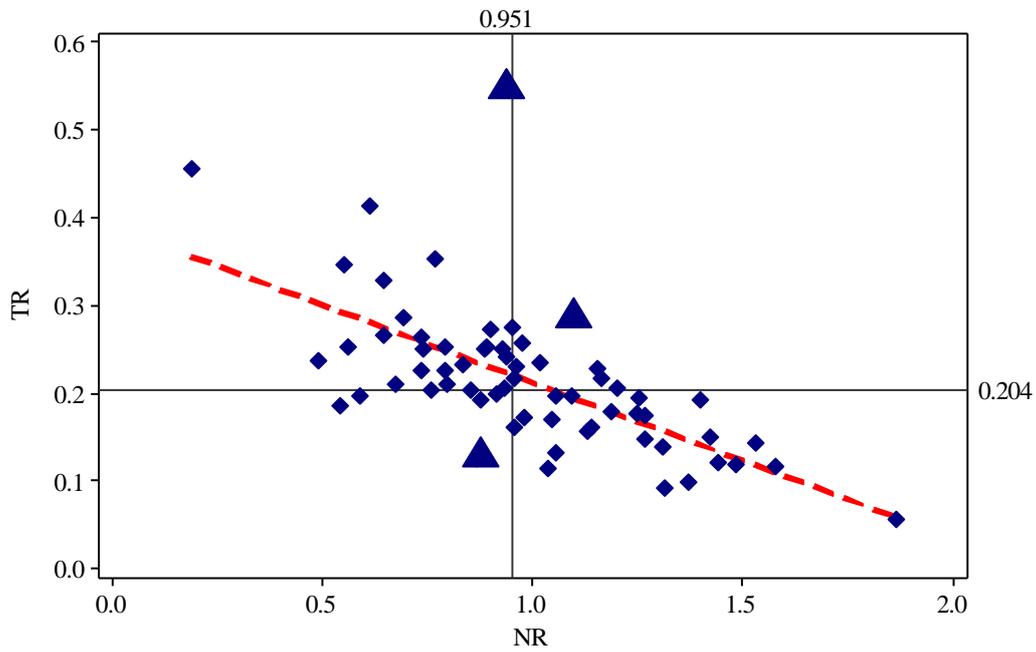

*Figure 4: Scatter plot of NR versus TR: references lines are the medians of the distributions for each variable; triangles indicate universities that stand out from the general trend line*

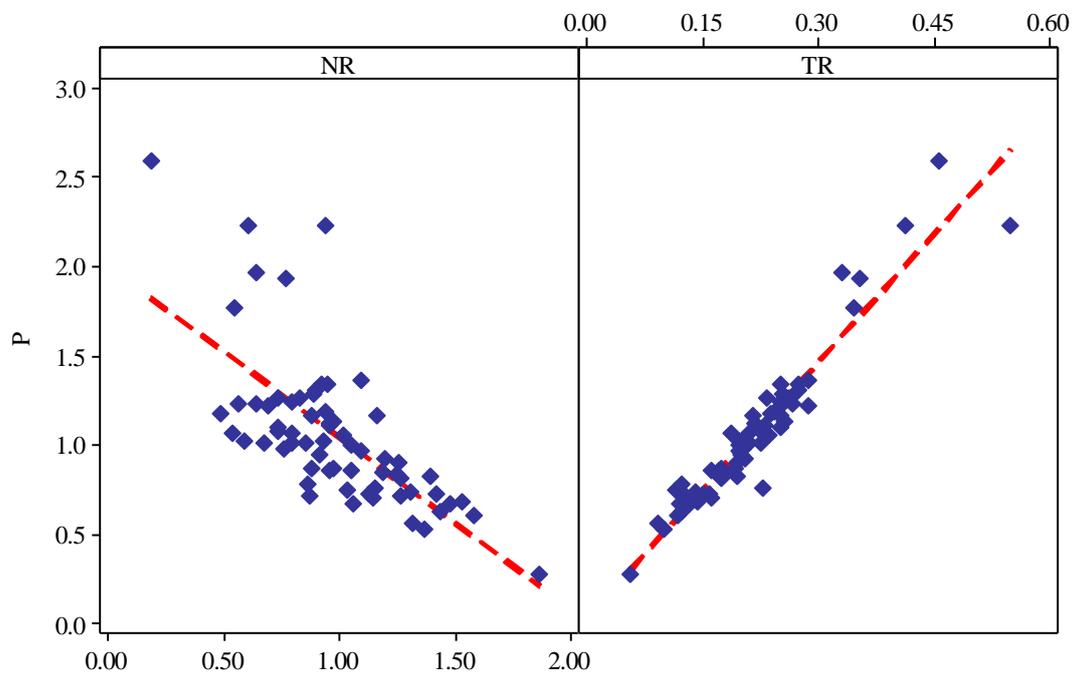

*Figure 5: Scatter plots of indices of unproductive (NR) and top (TR) professors vs university productivity (P)*

Given these results, we conclude our analyses by attempting a simple regression model to evaluate the effect of unproductive and top professors on the overall productivity of the university (*P*). To do this, we create a dummy variable to distinguish public universities (dummy=1) from private universities (dummy=0). The model results are shown in Table 6.

| Predictor | Coeff. | SE Coeff. | t | p-value | 95% Conf. Interval | | Beta | VIF |
|---|---|---|---|---|---|---|---|---|
| Constant | 0.518 | 0.110 | 4.72 | 0.000 | 0.298 | 0.737 | | |
| NR | -0.203 | 0.060 | -3.44 | 0.001 | -0.320 | -0.848 | -0.108 | 1.68 |
| $TR_{100}$ | 0.042 | 0.002 | 19.68 | 0.000 | 0.038 | 0.046 | 0.843 | 1.72 |
| Dummy | -0.167 | 0.051 | -3.24 | 0.002 | -0.271 | -0.064 | -0.145 | 1.03 |
| Post estimation statistics | | | Value | p-value | | | | |
| R squared | | | 0.935 | | | | | |
| R squared (adjusted) | | | 0.932 | | | | | |
| F test | | | 292.25 | 0.000 | | | | |
| Normality test of residuals* | | | 0.771 | 0.043 | | | | |

*Table 6: Linear regression analysis with 65 universities*
*Anderson-Darling test of normality

The regression equation obtained is:

$$P = 0.518 - 0.203 \cdot NR + 0.042 \cdot TR100 - 0.167 \cdot dummy + \varepsilon.$$

where the value of TR has been multiplied by 100. The regression coefficients are all jointly significant (F test=292.25 with a p-value of 0.000). The VIF test confirms the absence of multi-collinearity, which would be expected given the value of near 0.7 for correlation of TR with NR. Normality test of errors has a p-value around 5% (0.043), showing uncertainty regarding the effective normality of the regression residues. Still, the model explains more than 90% of total variance (R squared=0.935), which indicates a good fit.

For every unit increase in TR percentage, a 0.042 unit increase in global performance P is predicted, holding the NR constant; while for every unit increase in NR, a -0.203 unit decrease in global performance P is predicted. With the different units of measure for TR, expressed as a percentage, and NR, which is a non-dimensional indicator, it is naturally impossible to directly compare the two regression coefficients and give an appropriate interpretation of a unit increase in the NR variable. For this reason, we also calculate the beta standardized coefficients, which show that a percentage increase in TR has a higher effect on productivity than an equal decrease in NR.

The constant variable is 0.518 when the university is private, 0.351 when public; thus when a university (either public or private) does not have either top or unproductive professors it will place in the last quartile for performance P, with upper limit at 0.772. Meanwhile, for a public university with an unproductive professors index value equal to 1, the minimum concentration of top professors to achieve productivity greater than the median is 21%, and a concentration of 26% would place the university in the top performance quartile (lower limit 1.230).

## 4. Concluding remarks

In competitive higher education systems, such as "Anglo-Saxon" examples, we witness the concentration of top professors in elite universities and low performers in lower tier universities. In non-competitive higher education systems, such as the one in Italy, the distribution of both top and low performers is relatively uniform among universities, such that the variability of research performance is much higher within universities than between universities (Abramo et al., 2012a). In these systems, the adoption of selective funding policies based on the results of national research assessment exercises, could result as inequitable and counter-productive as demonstrated by Abramo et al. (2011b). In fact, top performers in low-tier universities could receive fewer funds than low performers who happen to be employed in high-tier universities. In Italy, with the particular context strongly structured against forced departure from public employment (it is unheard of to suspend tenure for low productivity), and with an absence of incentive systems, where faculty salaries are independent of productivity and linked only to academic rank and seniority, the situation is one where low-tier universities can only hope to improve their performance over the long term, through generational change. However the "selective" reduction of government funding could negate even this hope and also undermine the basic welfare principle of the Italian academic system: all students must be guaranteed opportunity of access to equal quality university education, independent of personal standing or geographic location. In such contexts, it would thus be preferable to allocate selective government funding directly to individual professors. In this regard, the difficulty of implementing performance measurement systems robust and reliable at individual level cannot be invoked for not to proceed in this direction. Recent advancements in bibliometric techniques, at least in the hard sciences, make possible the development of effective decision support systems based on large-scale measurement of bibliometric performance of individual researchers, offered as an aid for resource allocation and strategic planning in public research organizations (Abramo and D'Angelo, 2011).

In this study we have demonstrated that the possibility of calculating university performance net of unproductive professors, for potential allocation of resources based on such a ranking list, would not lead to notable variations in rank. Rather it is the top performers who have a more relevant impact on university performance. This leads to two fundamental recommendations, which sound obvious in efficient systems: a) for the policy maker, to consider funding allocations based on individual level rankings rather than on university rankings, in order to support maximal productivity among the research community; b) for universities, to pay particular attention to recruitment. Unfortunately, Italian legislation, for the moment, still requires that all faculty employment take place through national competition exams, often invalidated by diffuse practices of favoritism. In such non-competitive systems, it could be considered a more daring policy aimed at encouraging the budding of few spin-off universities made of the top scientists from the incumbents. These universities would by nature be highly immune to the favoritism virus and much more inclined to adopt practical and principled strategies, typical of those in competitive systems.

# References

Abramo, G., Cicero, T., D'Angelo, C.A. (2012a). The dispersion of research performance within and between universities as a potential indicator of the competitive intensity in higher education systems. *Journal of Informetrics,* 6(2), 155-168.

Abramo, G., Cicero, T., D'Angelo, C.A. (2012b). Revisiting size effects in higher education research productivity. *Higher Education*, 63(6), 701-717.

Abramo, G., D'Angelo, C.A., Di Costa, F. (2011a). Research productivity: are higher academic ranks more productive than lower ones? *Scientometrics*, 88(3), 915-928.

Abramo, G., Cicero, T., D'Angelo C.A. (2011b). The dangers of performance-based research funding in non-competitive higher education systems. *Scientometrics*, 87(3), 641-654.

Abramo, G., D'Angelo, C.A. (2011). National-scale research performance assessment at the individual level. *Scientometrics*, 86(2), 347-364.

Abramo, G., D'Angelo, C. A., Di Costa, F. (2008). Assessment of sectoral aggregation distortion in research productivity measurements. *Research Evaluation,* 17(2), 111-121.

D'Angelo, C.A., Giuffrida, C., Abramo, G. (2011). A heuristic approach to author name disambiguation in bibliometrics databases for large-scale research assessments. *Journal of the American Society for Information Science and Technology,* 62(2), 257-269.

Halffman, W., Leydesdorff, L. (2010). Is inequality among universities increasing? Gini coefficients and the elusive rise of elite universities. *Minerva,* 48, 55-72.

Hicks, D. (2012). Performance-based university research funding systems. *Research Policy*, 41(2), 251-261.

Horta, H., Huisman, J., Heitor, M. (2008). Does competitive research funding encourage diversity in higher education? *Science and Public Policy,* 35(3), 146-158.

Leydesdorff, L. (2008). Caveats for the use of citation indicators in research and journal evaluation. *Journal of the American Society for Information Science and Technology,* 59(2), 278–287.

Lundberg, J. (2007). Lifting the crown-citation z-score. *Journal of Informetrics*, 1(2), 145–154.

Moed, H.F. (2005). Citation Analysis in Research Evaluation. Springer, Berlin/Heidelberg/New York. ISBN: 978-1-4020-3713-9.

Osterloh, M., Frey, B.S. (2008). Creativity and Conformity: Alternatives to the Present Peer Review System. Paper presented at the workshop on *Peer Review Reviewed*. Berlin: WZB, 24-25 April 2008.

QS-Quacquarelli Symonds (2011). *World University Rankings*. Available at http://www.topuniversities.com/university-rankings/world-university-rankings/ Last access July 21, 2012.

SJTU-Shanghai Jiao Tong University (2011). *Academic Ranking of World Universities.* Available at http://www.shanghairanking.com/ARWU2011.html. Last access July 21, 2012.

THES-Times Higher Education Supplement (2012). *World Academic Ranking 2011-2012*. Available at http://www.timeshighereducation.co.uk/world-university-rankings/2011-2012/top-400.html. Last access July 21, 2012.

van der Ploeg, F., Veugelers, R. (2008). Towards evidence-based reform of European universities. *CESifo Economic Studies*, 54(2), 99-120.